\documentclass[aps,floatfix,prd]{revtex4}

\usepackage{graphicx}
\usepackage{dcolumn}
\usepackage{bm}
\usepackage{amsmath}
\usepackage{color}
\begin{document}
\def\dq{\frac{d^3q}{(2\pi)^3}\,} 
\def\be{\begin{equation}}
\def\ee{\end{equation}}
\def\X{$X$(2900)~}
\def\b{\bar b}
\def\q{\bar q}
\def\dd{\bar d}
\def\u{\bar u}
\def\s{\bar s}
\def\K{\bar K}
\def\Q{\bar Q}
\def\*{^{(*)}}
\def\bit{\begin{itemize}}
\def\eit{\end{itemize}}
\def\MeV{\textrm{ MeV}}
\def\nb#1{{\color{red}#1}}
\def\blue#1{{\color{blue}#1}}

\title{Kinematical Cusp and Resonance Interpretations of the $X(2900)$}

\author{T.J. Burns}
\affiliation{Department of Physics, Swansea University, Singleton Park, Swansea, SA2 8PP, UK.}
\author{E.S. Swanson}
\affiliation{
Department of Physics and Astronomy,
University of Pittsburgh,
Pittsburgh, PA 15260,
USA.}

\date{\today}

\begin{abstract}
We examine whether the LHCb vector $ud\bar{c}\bar{s}$ state $X(2900)$ can be interpreted as a kinematical cusp effect arising from $\bar D^*K^*$ and $\bar D_1 K\*$ interactions. The production amplitude is modelled as a triangle diagram with hadronic final state interactions. A satisfactory fit to the Dalitz plot projection is obtained that leverages the singularities of the production diagram without the need for  $\bar{D}K$ resonances. A somewhat  better fit is obtained if the final state interactions are strong enough to generate resonances, although the evidence in favour of this scenario is not conclusive.  

\end{abstract}

\maketitle 

\section{Introduction}

The LHCb collaboration has announced the discovery of a $\bar D K$ enhancement in the reaction $B\to D\bar D K$ that can be interpreted as Breit-Wigner resonances with parameters~\cite{LHCbX}:
\begin{align}
X_0(2866)&;\, J^P = 0^+, \quad &&M = 2866.3 \pm 6.5 \pm 2 \MeV,\ &&\Gamma = 57.2 \pm 12.2 \pm 4.1\MeV,\ &&\textrm{fit fraction} = 6\%,\\
X_1(2904)&;\, J^P =1^-, \quad &&M = 2904.1 \pm 4.8 \pm 1.3  \MeV,\ &&\Gamma = 110.3 \pm 10.7 \pm 4.3 \MeV,\  &&\textrm{fit fraction} = 31\%.
\end{align}
The discovery adds to a burgeoning list of putative hadrons of unconventional structure that have revitalised hadronic spectroscopy and shed light on the dynamics of QCD in the nonperturbative regime~\cite{Lebed:2016hpi}. 
 
The $X(2900)$ pair is observed as an enhancement in $D^-K^+$ and is therefore manifestly flavour-exotic, with a minimal quark content of  $ud\bar{c}\bar{s}$. It is therefore essential to determine the origin of the structure, for example it could be a novel  $\bar D^{(*)}K^{(*)}$ (or related) bound state, a compact tetraquark, or a production mechanism effect. Discriminating among interpretations would provide important diagnostic information on the workings of nonperturbative QCD. 

In this regard, the related $DK$ system has already provided evidence of `unconventional' dynamics since it is suspected that it strongly influences the enigmatic $D_{s0}(2317)$. An early argument for a $DK$ molecular interpretation is in Ref.~\cite{Barnes:2003dj}. More recently, a lattice gauge computation finds evidence that the $D_{s0}(2317)$ is an isoscalar $DK$ bound state, with hints of $J^P = 1^+$ and $2^+$ states at higher energy~\cite{Cheung:2020mql}. A review can be found in Ref.~\cite{Swanson:2006st}. Note, however, that the analogy between the $DK$ and $\bar DK$ systems may be misleading. The $DK$ system couples via the strong interaction to $c\bar s$, and this is presumably important for $D_{s0}(2317)$ and related states. This coupling is absent for the $\bar DK$ system.

Other flavour-exotic systems have been and gone. The $X(5568)$ claimed by the D0 experiment \cite{D0:2016mwd} would have flavour content $su\bar b\bar d$. We and others (for example \cite{Burns:2016gvy,Guo:2016nhb}) pointed out the implausibility of explaining the state using any of the standard approaches. Subsequently a number of experiments searched for, but did not find, the state~ \cite{Aaij:2016iev,Sirunyan:2017ofq,Aaltonen:2017voc,Aaboud:2018hgx}.

A variety of techniques and suggestions concerning the $X(2900)$ pair have appeared recently. Amongst these are the interpretation of the $J^P=0^+$ enhancement as an isosinglet compact tetraquark that is an analogue of the anticipated $ud\bar b\bar b$ tetraquark~\cite{Karliner:2020vsi}. A similar conclusion was reached in Ref.~\cite{He:2020jna}, where it was argued that the scalar state is a radially excited tetraquark while the vector state is an orbitally excited tetraquark. Both of these references used effective models with spatially constant interactions to reach their conclusions. Similarly, the author of Ref.~\cite {Zhang:2020oze} finds that a QCD sum rule computation with a scalar-scalar current supports the identification of the scalar signal with a $0^+$ tetraquark. An alternative approach is used in Ref.~\cite{Lu:2020qmp}, where a constituent quark model is solved variationally in a Gaussian basis. The authors find an extensive $ud\bar{s}\bar{c}$ spectrum, but conclude that the mass spectrum (including four scalars at 2765, 3065, 3152, and 3396~MeV) cannot accommodate $X(2900)$.

The tetraquark interpretation for $ud\bar s\bar c$ is motivated in part by the analogy with $ud\bar b\bar b$, for which there is evidence from lattice QCD for a bound state \cite{Bicudo:2015kna,Bicudo:2016ooe,Leskovec:2019ioa,Junnarkar:2018twb,Francis:2016hui}. The analogy is questionable considering that the \X states, being far above threshold, are not bound states. In this context it is noteworthy (and has hardly been discussed in recent literature) that the same lattice QCD calculation which finds a bound $bb\bar u\bar d$ system does {not} find binding in $ud\bar s\bar c$  \cite{Hudspith:2020tdf}.  Similarly there is no evidence for bound $ud\bar s\bar c$ states in early quark model calculations
\cite{Zouzou:1986qh}, or a recent QCD sum rules study \cite{Agaev:2019wkk}. If the \X states have a tetraquark nature, they may be very different from their bound $ud\bar b \bar b$ analogues.

In the molecular model, we note an early prediction for an isoscalar $\bar D^*K^*$ state, with $J^P=0^+$, and with mass and decay width comparable to the observed $X_0$ \cite{Molina:2010tx}. The model, which is based on the vector hidden gauge formalism, was recently refined in response to the LHCb discovery, and the authors predict partner states with $J^P=1^+$ and $2^+$, but do not propose an interpretation of the observed $1^-$ state $X_1$ \cite{Molina:2020hde}. Liu \textit{et al.~}use an effective Lagrangian that couples heavy quark fields and light mesons to compute binding energies of possible $\bar{D}K$, $\bar{D}^*K$,  $\bar{D}K^*$ and $\bar D^*K^*$ molecules~\cite{Liu:2020nil}. They argue that $X_0$ can be interpreted as {an isoscalar} $\bar{D}^*K^*$ molecule, but find no viable explanation for $X_1$. By contrast He and Chen \cite{He:2020btl}, with a similar model, reach a different conclusion, favouring an isovector interpretation for the $X_0$ as a $\bar D^*K^*$ molecule, and proposing that $X_1$ is a virtual state -- also isovector -- from the $\bar D_1 K$ interaction. Hu \textit{et~al.~}\cite{Hu:2020mxp} argue that $X_0$ is an isoscalar $\bar D^* K^*$ molecule and, using heavy-quark symmetry, they predict several partner states.

Other papers consider the possible role for both molecular and diquark degrees of freedom. Refs.~\cite{Chen:2020aos,Xue:2020vtq} both advocate $X_0$ as an isoscalar $\bar D^* K^*$ molecule, and whereas  Ref.~\cite{Xue:2020vtq} does not propose an explanation for $X_1$, Ref.~\cite{Chen:2020aos} argues that  $X_1$ is a P-wave diquark-antidiquark state.
Finally, production and decay of the $X(2900)$ pair was studied in Ref.~\cite{Qin:2020zlg}.

We do not offer a detailed analysis of these recent results; we note, rather, that many of the conclusions are based on methods that have not been carefully validated in the multiquark sector. It is therefore important to continue to explore options for the LHCb signal. Here we examine the possibility that a kinematical cusp can give rise to enhancement in $\bar{D}K$. Cusps have been invoked as explanatory mechanisms in hadronic physics for many years, most prominently by Bugg, who used them to describe the $Z(4430)$~\cite{Bugg:2007vp}, threshold synchronisation~\cite{Bugg:2008wu}, and the $Z_b(10610)$ and $Z_b(10650)$~\cite{Bugg:2011jr}. More recently, there has been a surge of interest in the explanatory power of triangle and other singularities, with many applications~\cite{ghlmz, clm, Swanson:2014tra, Swanson:2015bsa, Guo:2015umn, Liu:2015fea, aps, Ikeda:2016zwx}. 
We remark that we do not distinguish two-body threshold (Wigner) cusps from triangle singularities in the following as these are intertwined in the production mechanism; rather we refer to any enhancement that appears due to the production portion of the process as a kinematical cusp. These concepts are usefully reviewed in Ref.~\cite{Guo:2019twa}.

We explore the implications of a triangle diagram coupling between an initial $B^+$ meson and the final $D\bar{D}K$ state. As we have noted, such a production mechanism can naturally lead to enhancements in rates at thresholds that correspond to nearly on-shell intermediate states. In this case $\bar{D}^*K^*$ and $\bar D_1(2420)K$ thresholds occur at 2902 and 2917~MeV respectively, and so offer a promising mechanism for investigation. (We will from now on use $\bar D_1$ to stand for $\bar D_1(2420)$; the partner state $\bar D_1(2430)$ is significantly broader and is thus unlikely to have a role in the formation of the narrow $X(2900)$ structures.)

The question naturally arises as to why structures would appear near to these particular thresholds, and not at the thresholds for other combinations of related hadrons. Here the assumed dominance of one-pion exchange gives some insight. It is natural for structures to appear near the  $\bar D^* K^*$ threshold, since this is the lightest combination of hadrons with this flavour that can interact via elastic one-pion exchange. There is no elastic one-pion exchange for $\bar D K$, $\bar D K^*$, or $\bar D^* K$, and the large energy gaps between thresholds suppress the importance of inelastic transitions.

The role of $\bar D_1 K$ also emerges naturally with this kind of argument. Although elastic $\bar D_1 K$ scattering is not allowed via one-pion exchange, the inelastic transition  $\bar D_1 K\to\bar D^*K^*$ is allowed, and the proximity of the $\bar D_1 K$ and $\bar D^*K^*$ thresholds implies that this coupling could be important. Indeed we have shown that an analogous coupling $\Lambda_c(2595)\bar D\to \Sigma_c\bar D^*$, where again the thresholds are almost degenerate, gives rise to an attractive potential which may explain the $P_c(4457)$ state~\cite{Burns:2015dwa,Geng:2017hxc,Burns:2019iih}. The unusual feature common to both of these scenarios is the presence of a pion vertex coupling a hadronic ground state to an orbital excitation.

Out of all of the possible channels with $\bar D K$ flavour, pion exchange points to the particular significance of the $\bar D^*K^*$ and $\bar D_1 K$ channels, so it is therefore very striking that it is precisely at their thresholds where the \X enhancements appear.

We also remark that molecular models with only $\bar  D^*K^*$ degrees of freedom struggle to explain the  $1^-$ quantum numbers of the heavier $X_1(2904)$, as they imply the constituents are in a relative P-wave. With the $\bar D_1 K$ channel, however, the $1^-$ quantum numbers are very natural, as here the constituents can be in relative S-wave.

While preparing this report, we became aware of Ref.~\cite{Liu:2020orv} (see also Ref.~\cite{Huang:2020ptc}) that examines a very similar triangle mechanism. Their scope is restricted to noting the expected enhancement in the $\bar{D}K$ invariant mass distribution at approximately 2900 MeV, and to substantial broadening of the peak when realistic $K^*$ and $D_1$ widths are taken into account. As described in the next section, we go beyond these initial observations by making a detailed fit to the mass distribution, by incorporating nonperturbative final state interactions (FSI), and by successfully allowing for the measured $K^*$ and $D_1$ widths. We find that it is possible to fit the mass distribution very well with the $\bar{D}^*K^*-\bar D_1K$ cusp effect. An even better fit is obtained when $\bar D_1K^*$ and strong final state interactions are considered.

\section{Triangle Final State Interaction Models}

\subsection{Production}

The LHCb collaboration found that the angular distribution (in $\bar{D}K$) provides strong evidence for a $1^-$ Breit-Wigner component in their amplitude model. The fit was further improved by including the $X_0(2866)$ Breit-Wigner, although we note that its contribution to the angular distribution is negligible. As such, we suspect that the effect of the $X_0$ can be mimicked by background or other dynamics and that it will prove to be unnecessary.  Because of this we focus attention on the vector $X_1(2904)$ in the following. 

\begin{figure}[ht]
\includegraphics[width=6cm,angle=90]{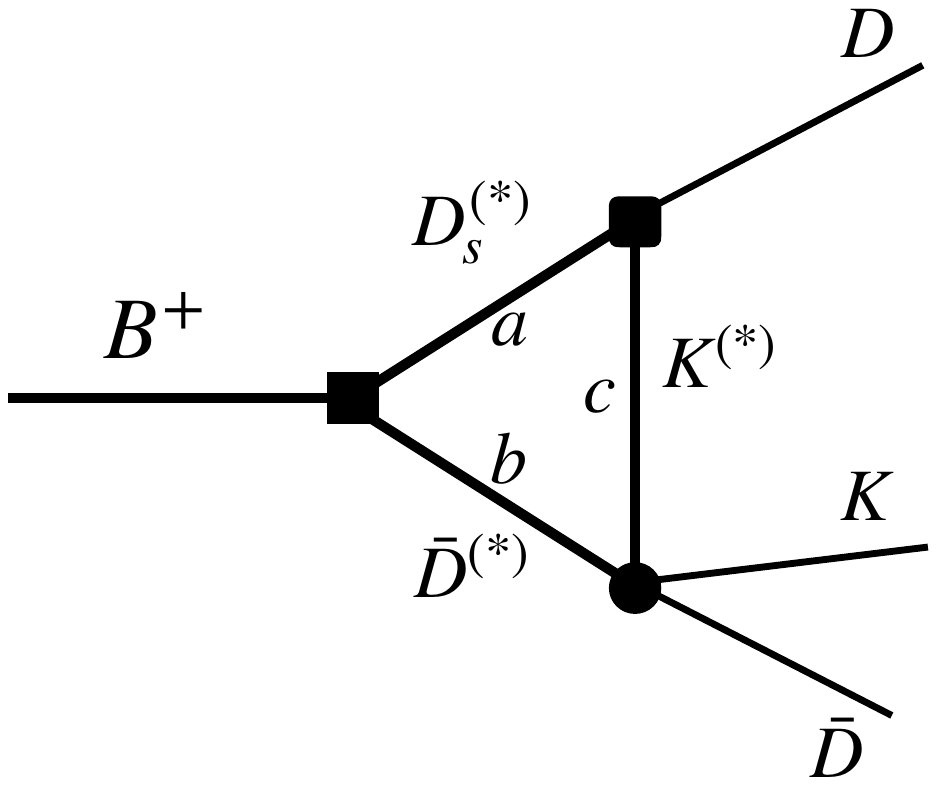}
\caption{Production Model for the $X(2900)$.}
\label{fig:triangle}
\end{figure}

The process we consider is illustrated in Fig.~\ref{fig:triangle}. In general we consider three intermediate mesons labelled $a$, $b$, and $c$. Meson combinations that dominate the process should be (i) colour enhanced, (ii) S-wave where possible, and (iii) permit the triangle to go nearly on-shell. Possible combinations are shown as $D_s^{(*)} \bar{D}^{(*)} K^{(*)}$ in the figure.  Experimentally, the largest branching ratios of the $B^+$ are to states accessible via colour-enhanced transitions such as $\bar{D}_s^*D$ (at order one percent) so we focus on these. However, one expects a series of ``$D_s$" states to contribute, with those near $m(B^+) = m(D_s) + m(D)$ dominating the sum. We account for this by using an effective $D_s^*$ meson of spin-parity $1^-$ and mass of 3 GeV in the following. (We will also examine the effect of changing this mass.)

We note in passing that the model of Ref.~\cite{Liu:2020orv} for $X_0(2866)$ relies on intermediate states that arise only from colour-suppressed transitions, specifically  $\chi_{c1} K^*$, where $\chi_{c1}$ is either $\chi_{c1}(3872)$, $\chi_{c1}(4140)$ or $\chi_{c1}(4274)$. It is questionable whether colour-suppressed transitions could produce prominent kinematical enhancements over a background of colour-enhanced processes. For comparison we note that, experimentally, $\mathcal B(B^+\to X(3872)K^+)<2.6\times 10^{-4}$, whereas colour-favoured transitions are typically at the percent level.

As indicated in Fig.~\ref{fig:triangle}, mesons $b$ and $c$ will be taken to have flavour $\bar{D}K$.  Clearly, choosing $bc = \bar{D}^*K^*$ and $\bar D_1K$ will be important to creating reaction strength near 2900 MeV. For completeness we shall expand this collection to include related channels which can also couple to $1^-$, specifically $\bar{D}K|_P$, $\bar{D}^*K^*|_P$, $\bar D_1K|_S$, and $\bar D_1K^*|_S$, where the partial wave is indicated with a subscript. (We do not include $\bar{D}^*K|_P$, $\bar{D}K^*|_P$ for reasons discussed later.) In practice we will find that an excellent fit to the experimental data can be obtained with only a subset of these channels.

The next step is to model the vertices in the triangle diagram. The electroweak vertex will be written in terms of the heavy quark effective field, with the transition $B\to D^{(*)}$ represented by the Isgur-Wise function $\xi(w)$ with $\xi \sim (2/(1+w))^2$ and the orbital transition $B\to D_1$ represented by the Isgur-Wise function $\tau(w)$. The variable $w$ is defined as $p_B\cdot p_D/(m_B m_D)$ and the function $\tau$ will be approximated by $(2/(1+w))^6$\cite{Ebert:1999ga}. Expressing $w$ in terms of three momenta is frame-dependent\cite{Lakhina:2006vg}; we make this mapping in the $B$ meson rest frame. Decay constants, CKM matrix elements, etc will be absorbed into coupling strengths (to be defined shortly) and will henceforth be suppressed. We shall see that dependence on the form of the electroweak vertex can be readily absorbed into model coupling parameters.

The upper vertex represents a strong decay and we have chosen to model it with the well-known `3P0' model of hadronic transitions~\cite{3p0}. Thus, for example, the transition $D_s^* \to DK$ is approximately proportional to $x\exp(-x^2/12)$, where $x=p/\beta_\textrm{3P0}$, $p$ is the relative final state momentum, and $\beta_\textrm{3P0}$ is a universal width parameter describing simple harmonic oscillator wavefunctions for the mesons. Although this form is typical, we believe that the exponential decay is not physical for large momentum and have replaced this functional form with 
$$
\frac{x}{1+x^2/12}
$$
though, in practice, this makes little difference.  Once again, the strength of the vertex will be absorbed into a global coupling.

The remaining vertex (the solid circle in Fig. \ref{fig:triangle}) represents the final state interactions in the $\bar{D}K$ system, which will be discussed in the next section. For now we denote the relevant portion as $F_L\cdot Y_{LM}$. 
Finally, we use nonrelativistic kinematics as all the mesons are predominantly moving slowly in the $B^+$ rest frame. The resulting expression for the triangle diagram is

\begin{eqnarray}
\triangle_{a\alpha}(s_{\bar{D}K}) &=& \int \dq F_\textrm{ew}(\bm{q}+\bm{k}/2)\, F_\textrm{3P0}(3\bm{k}/4 - \bm{q}/2) \, F_{L_\alpha}(q) Y_{L_\alpha M_\alpha}(\hat q) \cdot \nonumber \\
&& [m_B - m^\alpha_a - m^\alpha_b - (\bm{q}+\bm{k}/2)^2/(2\mu_{ab}) + i \Gamma^\alpha_{a}/2 + i\Gamma^\alpha_b/2]^{-1} \cdot \nonumber \\
&& [m_B-E_D - m_b^\alpha - m_c^\alpha - (\bm{q}+\bm{k}/2)^2/(2 m_b^\alpha) - (\bm{q}-\bm{k}/2)^2/(2 m_c^\alpha) + i \Gamma^\alpha_{b}/2 + i \Gamma^\alpha_c/2]^{-1}.
\label{eq:tri}
\end{eqnarray}
The notation ``$a\alpha$" indicates that the diagram depends on the meson $a$ and the channel index $\alpha$, which contains a meson pair $bc$. We have allowed for an angular momentum $L_\alpha$ (normally 0 or 1) in the final state interaction vertex. Widths of the mesons $a$, $b$, and $c$ are also accounted for as they can have a strong attenuation effect on the amplitude. Finally, $\bm{k}$ is the momentum of the outgoing $D$ meson while its energy is $E_D$; these quantities can be used to obtain the Mandelstam variable, $s_{\bar{D}K}$.

\subsection{Final State Interactions}

We have argued that four channels are naturally relevant to the description of $\bar{D}K$ production in $B$ decay and have chosen to use one-pion-exchange as a guide in modelling the final state interactions among these channels. Thus interactions that carry $D\to D$, $K\to K$, and $D_1 \to D$ are disallowed.

Although this description of the final state interactions can be carried to completion, we find it expedient to simplify and generalize it somewhat. Thus we replace the central, tensor, and vector interactions of the one-pion-exchange scenario with regulated separable interactions that are described by universal form factors and channel couplings, as follows:
\begin{equation}
\langle p L M; \alpha | V | p' L' M'; \alpha'\rangle = \lambda_{\alpha\alpha'} Y_{LM}(\hat p) F_L(p) \cdot Y^*_{L'M'} (\hat p') F_{L'}(p').
\label{eq:V}
\end{equation}
The hadronic form factor is modelled as
\begin{equation}
F_L(x) = \frac{x^L}{1+x^2}, \ \  x = \frac{p}{\beta},
\end{equation}
where $\beta$ is a universal scale and the numerator implements the expected angular momentum barriers in hadronic decays and interactions.


In this way we arrive at the interaction potential matrix shown in Tab. \ref{tab:fsi2}. We have chosen to neglect the low mass $\bar{D}^*K$ and $\bar{D}K^*$ channels as these do not couple strongly to the rest of the system and are not expected to contribute strongly to the dynamics at 2900 MeV. Furthermore, we do not include $\bar{D}K$ as a driving (triangle) amplitude since this subamplitude is only expected to contribute at the left edge of phase space.

\begin{table}[ht]
\caption{Final State Interaction Model}
\begin{tabular}{c|cccc}
\hline\hline
$\bm{\lambda}(1^-)$ & $\bar{D}^*K^*|_P$  & $\bar D_1K|_S$ & $\bar D_1K^*|_S$ & $\bar{D}K|_P$ \\
\hline
$\bar{D}^*K^*|_P$ & $C_1$ & $C_2$ & $C_3$ & $C_4$ \\
$\bar D_1K|_S$   &       &  0    & $C_5$ & 0 \\
$\bar D_1K^*|_S$ &       &       & $C_6$ & 0 \\
$\bar{D}K|_P$     &       &       &       & 0 \\
\hline\hline
\end{tabular}
\label{tab:fsi2}
\end{table}

In defining the final state interaction we choose to use pion exchange as a guide to identitfy zeroes in the interaction matrix; the remaining interactions have been denoted with six couplings, as shown in Tab. \ref{tab:fsi2}. 
 Apart from $\bar D K$, all of the channels are production channels, and each of these has an associated coupling constant (denoted as $g_{a\alpha}$ below) that is fit to data. Since the number of non-zero entries in the FSI matrix is comparable to the number of fit parameters, relative numerical factors in the matrix turn out to be unimportant in the case of weak coupling.

The $\bar D^* K^*$ channel includes several sub-channels $^1P_1$, $^3P_1$ and $^5P_1$. A full treatment of the interaction potential due to pion-exchange requires all of these channels to be included. But since each additional channel comes with its own fit parameter (the production coupling constant), the inclusion of all these channels does not ultimately further constrain the results. Hence for the sake of simplicity, we consider a single $\bar D^* K^*\vert_P$ channel.
Table \ref{tab:fsi2} reveals the importance of the $\bar{D}^*K^*$ channel, as this is the only channel that couples directly to the $\bar{D}K$ discovery mode.

A variety of final state interaction coupling strengths will be examined in the following, with the conclusion that the particular choices for the couplings $C_i$ do not affect the final fit qualities very much (although they do change the physical interpretation). As mentioned already, this is likely because each channel (apart from $\bar D K$) is also a production channel and therefore has an amplitude coupling that can offset changes to the FSI potential, and because the kinematical singularities are capable of describing the data (as will be shortly demonstrated).
Because the fits are rather insensitive to the structure of the FSI matrix, for simplicity we set the couplings $C_i$ that correspond to closely related transitions to be equal. Hence we adopt $C_1=C_4$ for $\bar D\*K\*\to\bar D\*K\*$, $C_2=C_3$ for $\bar D^*K^*\to \bar D_1K\*$, and $C_5=C_6$ for $\bar D_1K\*\to \bar D_1K\*$.	
       A more rigorous approach to the FSI potential would be to derive the long-distance component from one-pion exchange, and parametrise the short-distance behaviour using contact terms that are fit to experimental data. Unfortunately the large number of contact terms makes this approach untenable: each family of transitions $\bar D\*K\*\to\bar D\*K\*$,  $\bar D^*K^*\to \bar D_1K\*$, and $\bar D_1K\*\to \bar D_1K\*$ has an independent set of contact terms, and there is no prospect that such a large number of terms can be constrained with the current experimental data. The situation is very different compared to the corresponding model for $P_c$ states, where heavy quark symmetry implies just two contact terms for all potentials among S-wave $\Sigma_c\*\bar D\*$ channels (see for example Refs. \cite{Liu:2019tjn,Valderrama:2019chc,Sakai:2019qph,Du:2019pij}). We also note that the experimental data for $X(2900)$ is much less constraining than in the case of the $P_c$ states, with just one prominent experimental peak compared to three.

Nonperturbative final state interactions are obtained by solving the Bethe-Heitler equation, $T=V+VGT$, using standard techniques. In this case a reduced T-matrix is employed that is defined via
\be
\langle \vec{p} L M; \alpha| T | \vec{p}' L' M'; \alpha'\rangle  \equiv Y_{LM}(\hat p) F_L(p) \cdot t_{\alpha \alpha'}(p,p') \cdot Y^*_{L'M'}(\hat p') F_{L'}(p').
\label{eq:t}
\ee
We are now prepared to write the amplitude model taking into account the triangle production diagram and the full final state interactions represented by Eqs. (\ref{eq:V}) and (\ref{eq:t}), and Table~\ref{tab:fsi2}. The result is 
\begin{equation}
{\cal A} = \frac{g_\textrm{bg}}{m_B^2} + \sum_{a\alpha(bc)} \frac{g_{a\alpha}}{m_B}\, \triangle_{a\alpha}(s_{DK}) \cdot  t_{\alpha:DK}(s_{DK}) \cdot Y^*_{L_f M_f}(\widehat{k_{KD}}) F_{L_f}(k_{KD}).
\end{equation}
Notice that different triangle diagrams can drive the coupled channel system and that these are given couplings $g_{a\alpha}$ which are to be fit to the data. A background scattering term, $g_\textrm{bg}$, has been included in the amplitude model. This term is taken to be a fixed complex constant, as we are -- perhaps foolishly -- heavily biased against overly elaborate background models. Following common experimental modelling procedure, the couplings will be taken to be complex constants in the following. This may appear suspect from a theoretical point of view because one can regard the amplitude model as a product of an effective field theory that should have real couplings -- imaginary pieces of the amplitude are generated by dynamics that give rise to cuts and other singularities. However, hadronic models necessarily neglect strong dynamics, such as additional channels. Good models will be able to absorb the effects of the neglected dynamics in the couplings, but these must now be complex.

\section{Fit Results}

Our objective is to fit the $\bar{D}K$ invariant mass distribution obtained by the LHCb collaboration~\cite{LHCbX}. We follow the collaboration's lead and restrict the projection to $m(D^+D^-) > 4.0$ GeV, thereby removing many charmonia/$D\bar{D}$ resonances, which simplifies the amplitude model.  Of course, fitting a single projection misses much of the information available in the full Dalitz plot -- eventually fitting the amplitude model to the full data set would be very interesting.

Our approach is to adopt reasonable values for the form factor parameters, specifically
$$
\beta_\textrm{3P0} = 500 \MeV, \qquad \beta = 700 \MeV.
$$
These values are chosen because they are typical of quark model computations and because they provide results that are commensurate with the experimental data.
Furthermore, we choose to fix the final state interaction model according to different criteria. We report on three models in the following. These models are meant to represent weak final state interactions, moderate interactions, and interactions just strong enough to generate resonances  as determined by examining the final state system (see Tab. \ref{tab:C}).

\begin{table}[ht]
\caption{Final State Interaction Cases, (GeV)$^{-2}$}
\begin{tabular}{c|cccccc}
\hline\hline
case & $C_1$ & $C_2$ & $C_3$ & $C_4$ & $C_5$ & $C_6$ \\
\hline
weak &  2 & 3 & 3 & 2 & -3 & -3 \\
moderate & 10 & 20 & 20 & 10 & -30 & -30 \\
strong & 10 & 20 & 20 & 10 & -50 & -50 \\
\hline\hline
\end{tabular}
\label{tab:C}
\end{table}
With this approach, the only free parameters are the $N-1$ couplings $g_{a\alpha}$
and the background term $g_\textrm{bg}$, giving a total of $2N-1$ parameters.

\begin{figure}[ht]
\includegraphics[width=14cm,angle=0]{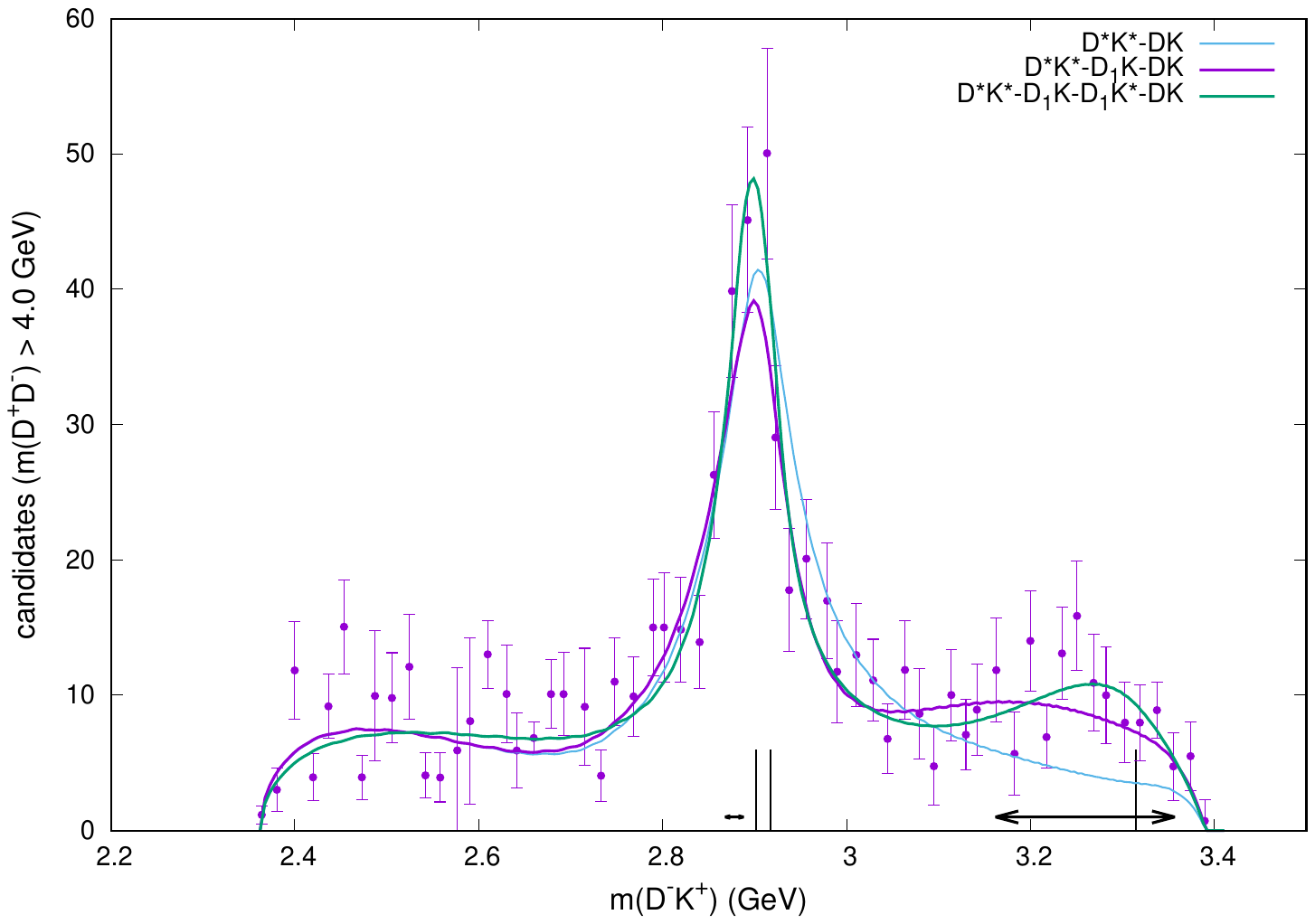}
\caption{LHCb invariant mass distribution and fit results. Vertical lines are thresholds $\bar{D}^*K^*$, $\bar D_1K$, and $\bar D_1K^*$. Horizontal lines represent pole positions in the strong FSI case.}
\label{fig:fits}
\end{figure}

In our experience it is too easy -- and far too common -- to fit an elaborate model to data and then ascribe physical reality to features of the model that are not warranted. We have therefore taken a conservative approach to the fit in which we progressively expand the model size, starting with the minimum number of channels required, $\bar{D}^*K^*$ and $\bar{D}K$. In this case the only relevant coupling is $C_1 = C_4$. The result of the three-parameter fit for the moderate and strong cases are shown as the blue line in Fig. \ref{fig:fits}; this fit has a chi-squared per degree of freedom of 1.98. It is evident that the $X(2900)$ peak is reasonably well described, while the chief failing of the model is in describing the high $m(\bar{D}K)$ distribution.

We next expand the model to three channels, $\bar{D}^*K^*$, $\bar D_1K$, and $\bar{D}K$, with the new channel selected because its threshold is close to that of $\bar{D}^*K$, and couples to it via one-pion exchange. In this case the relevant couplings are $C_1=C_4$ and $C_2=C_3$ and there are five parameters to be fit. The result for the moderate and strong cases is displayed as the purple line in Fig. \ref{fig:fits} and has $\chi^2/$dof = 1.36. Evidently the substantial improvement in the fit quality is due to the better description of the high mass region. 

The Dalitz plot reported in Ref.~\cite{LHCbX} contains a band of strength near $m(\bar{D}K) = 3.2$ GeV that appears as the weak enhancement seen near 3.2 GeV in Fig. \ref{fig:fits}. While this enhancement is not terribly significant, and it passed unmentioned in the LHCb analysis, it is too tempting for theorists to pass up. We therefore consider a four-channel model that adds the $\bar D_1K^*$ mode. This will naively add a triangle cusp 
near  $m(D_1) + m(K^*) = 3315$ MeV that could prove useful in describing the high mass region. The resulting fit for the strong FSI case is shown as the green line in Fig. \ref{fig:fits}. Once again, the fit quality has improved, with 
 $\chi^2$/dof = 1.22, largely due to a better fit through the high mass region\footnote{Parameter values in this case are 
$g_{D_s^{(*)}\bar{D}^*K^*} = 48.1$, 
$g_{D_s^{(*)}\bar{D_1}K} = -8.4 - i 46.4$, 
$g_{D_s^{(*)}\bar{D}_1K^*} = 20.9 + i 40.9$, 
and $g_\textrm{bg} = -51.1 + i 70.5$. }.

At this stage we choose to stop expanding the amplitude model because fits with more parameters will likely lose physical significance. Instead we proceed by analysing the fits that have been obtained to aid in their interpretation.
The first task is to examine model sensitivity by varying parameters and refitting for the four-channel case.
Since all the resulting fits are very similar to that shown in Fig. \ref{fig:fits}, we only report the chi-squared per degree of freedom as a measure of the resulting differences (in Table \ref{tab:refits}). The first three entries in the table refer to the FSI models of Table~\ref{tab:C}. We see that the strong case is marginally preferred. The remaining rows implement specific changes to the strong FSI or production models.  Some of the changes that are induced can be substantial; but, as the chi-squared shows, all changes can be almost completely absorbed into the free couplings of the amplitude model. As expected, changing the mass of meson $a$ (``$D_s^*$") as indicated in the final two rows has a substantial effect on the global strength of the amplitude; however, this change can be countered by a commensurate increase in the channel couplings. This is because the enhancements are dynamically generated in the strong coupling case. Alternatively, performing the same exercise in the two channel case reveals that the width of the peak near 2900 MeV is strongly dependent on the effective $D_s^*$ mass.

\begin{table}[ht]
\caption{Model Robustness Evaluation}
\begin{tabular}{ll}
\hline\hline
variation & $\chi^2$/dof \\
\hline
weak FSI                    & 1.36   \\
moderate FSI                & 1.44   \\  
strong FSI                  & 1.21   \\
$\beta = 500$ MeV           & 1.32   \\
replace EW vertex $\tau$ with $\xi$    & 1.22   \\  
$m($``$D_s^*$") = 2500 MeV  & 1.37   \\
$m($``$D_s^*$") = 2112 MeV  & 1.25   \\
\hline\hline
\end{tabular}
\label{tab:refits}
\end{table}

\begin{figure}[ht]
\includegraphics[width=12cm,angle=0]{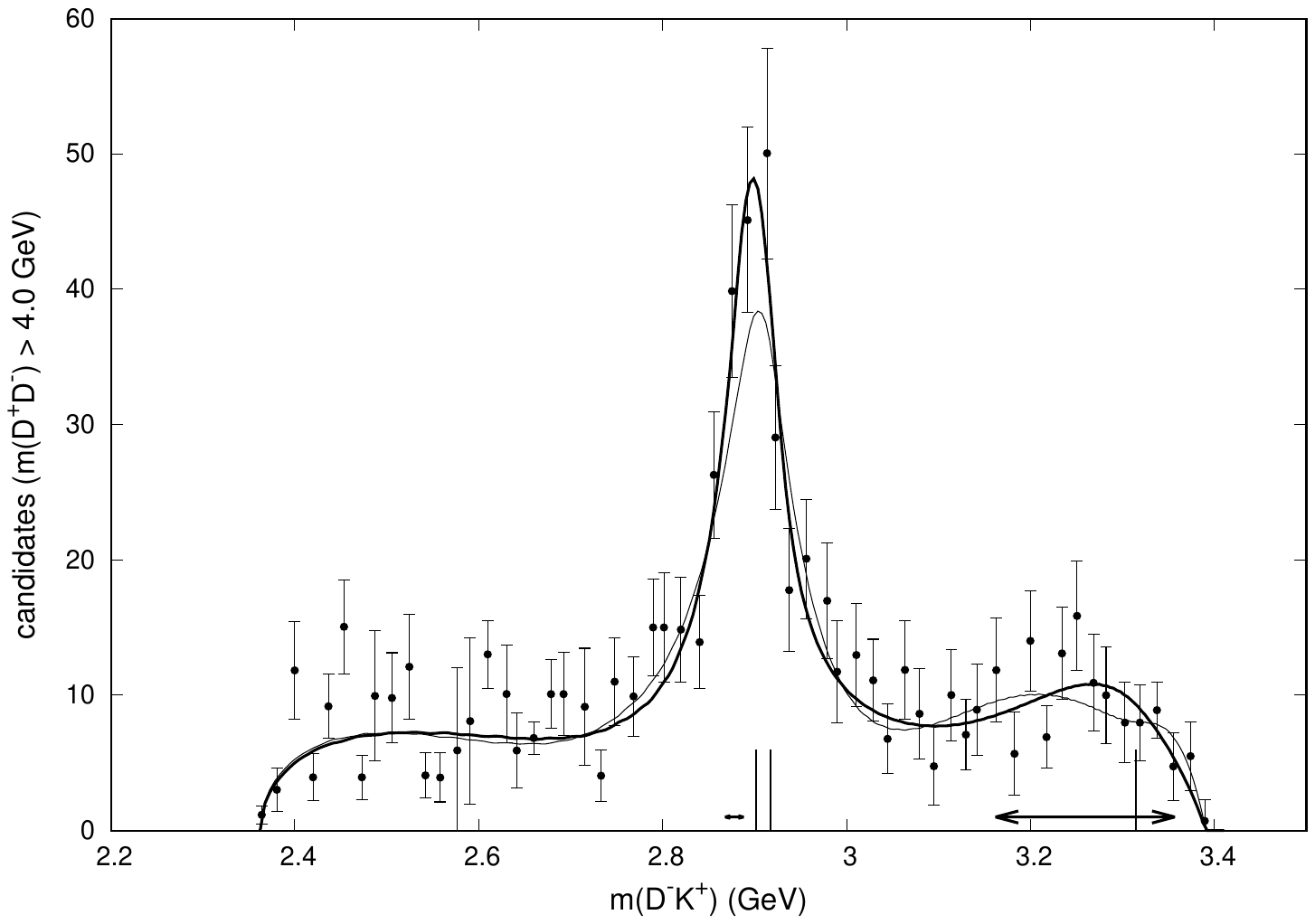}
\caption{LHCb invariant mass distribution and fit results. Solid curve: strong FSI case; thin curve: weak FSI case. Vertical lines are thresholds $\bar{D}^*K^*$, $\bar D_1K$, and $\bar D_1K^*$. Horizontal lines represent pole positions in the strong FSI case.}
\label{fig:fit4}
\end{figure}

We next turn attention to dependence on the assumed strength of the final state interactions. 
First we test dependence on the coupling  $C_1=C_4$ in the two-channel case. As expected, we find that this controls the width of the peak that is generated near 2900 MeV, with $C_1$ near 10 GeV$^{-2}$ creating peaks of approximately the correct shape.
We also refit the four-channel model in the weak FSI case, obtaining the results shown in Fig. \ref{fig:fit4}. With $\chi^2$/dof = 1.35, it is evident that the weak FSI case describes the data quite well, although it has a slightly poorer fit through the enhancement peak and the high mass region. Because the weak case generates little final state interactions, we conclude that kinematical cusps are capable of describing the $\bar{D}K$ mass distribution.

Alternatively, the interactions in the strong FSI case are sufficiently strong to generate resonances, which occur at 2878 - $i$20 MeV and 3260 - $i$195 MeV. These have structure dominated by $\bar{D}^*K^* - \bar D_1K$ and $\bar D_1K^*$ respectively. Since the strong FSI case has a better fit quality, these observations may be regarded as marginal evidence in favour of exotic (i.e., resonance-forming) dynamics in the $\bar{D}K$ system. Thus a more detailed fit to the full LHCb data set should help clarify the issue and would be interesting to pursue.

It is tempting to interpret the previous observations as evidence for exotic resonances, but as we have noted, non-exotic interpretations are also feasible. An indication of the lack of robustness of the exotic interpretation is obtained by setting $g_{D_s^*\bar{D}^*K^*}$ and  $g_{D_s^*\bar D_1K}$ to zero and replotting while leaving the other parameters fixed. The result is a very large amplitude, which must be cancelled against the other subamplitudes. This is an indication that the different subamplitudes are acting as a basis in the fit to the data, rather than physically motivated entities. In contrast, performing the same experiment in the moderate FSI case yields subamplitude of reasonable size that are readily interpretable. For this reason, we hesitate to make proclamations about the interpretation of the $X(2900)$, other than to note that simple kinematical cusps can explain the mass projection, and strong final state interactions can improve the fit.

\section{Conclusions}

A model of $D\bar{D}K$ production from $B^+$ decay has been developed that combines a triangle diagram production mechanism with nonperturbative final state interactions in the $\bar{D}K$ channel. The final state interactions were modelled with a separable potential with a structure motivated by one-pion-exchange phenomenology.  
We remark that the experimental widths of the $K^*$ and $D_1$ mesons have been used in the formalism. The reasonably narrow peak seen in Fig. \ref{fig:fits} is thus at odds with the broader peaks described in Ref.~\cite{Liu:2020orv}.

The results of a variety of fits to the $\bar{D}K$ invariant mass distribution yield strong evidence that the structure at 2900~MeV can be interpreted as a kinematical cusp due to the $\bar{D}^*K^*$ intermediate state, possibly enhanced by the $\bar D_1K$ mode. Unlike tetraquark models, this scenario does not imply a proliferation of partner states since, as we have argued, the $\bar{D}^*K^*$ and $\bar D_1K$ channels are uniquely important from the perspective of pion exchange.

Our model permits the generation of molecular states with flavour $ud\bar{c}\bar{s}$  and these are marginally preferred by the fits, with one such FSI generating poles at 2878 - $i$20 MeV and 3260 - $i$195. Nevertheless, we regard the evidence as weak, primarily because the strong FSI case yields large cancelling amplitudes, which raises the possibility that the good fit is due to intrinsic variability in the model as opposed to the quality of the underlying physics. Performing a fit to the full LHCb data set will provide vital further information that might be able to pin down the model characteristics.  It will also be interesting to describe the final state interactions with a one-pion-exchange formalism, as this will help to constrain the model. The question of the $X_0(2866)$ needs to be addressed: does the model developed here obviate the need for this component, or is it required by the full dataset?

Finally, we remark that other production and decay modes of the $X(2900)$ states are possible. The experimental  observation or otherwise of these modes can sharply discriminate among their possible interpretations. We explore these ideas in a forthcoming paper \cite{BS2}.

\acknowledgments

The authors are grateful to Daniel Johnson and Tim Gershon for discussions.
Swanson acknowledges support by the U.S. Department of Energy under contract DE-SC0019232.

\end{document}